\documentclass[prb,aps,twocolumn,superscriptaddress,lengthcheck,showpacs,amsmath,amssymb,floatfix,longbibliography]{revtex4-1}
\usepackage{graphicx}  
\usepackage{hyperref}
\usepackage{microtype}
\usepackage{float}
\usepackage{blindtext}
\usepackage{bbold}

\hypersetup{colorlinks,linkcolor=blue,citecolor=blue,urlcolor=blue}
\def\equationautorefname~#1\null{Eq.~(#1)\null}

\begin{document}

\title{Uncovering nonperturbative dynamics of the biased sub-Ohmic spin-boson model with variational Matrix Product States}

\author{C. Gonzalez-Ballestero}
\email{carlos.gonzalez-ballestero@uibk.ac.at}
\affiliation{Institute for Quantum Optics and Quantum Information of the Austrian Academy of Sciences, A-6020 Innsbruck, Austria}
\affiliation{Institute for Theoretical Physics, University of Innsbruck, A-6020 Innsbruck, Austria}
\affiliation{Departamento de F\'isica Te\'orica de la Materia Condensada and Condensed Matter Physics Center (IFIMAC), Universidad Aut\'onoma de Madrid, E-28049 Madrid, Spain}

\author{Florian A. Y. N. Schr\"oder}
\affiliation{Cavendish Laboratory, University of Cambridge, J. J. Thomson Avenue, Cambridge, CB3 0HE, UK}

\author{Alex W. Chin}
\affiliation{Cavendish Laboratory, University of Cambridge, J. J. Thomson Avenue, Cambridge, CB3 0HE, UK}


\begin{abstract}
We study the dynamics of the biased sub-Ohmic spin-boson model by means of a Time-Dependent Variational Matrix Product State (TDVMPS) algorithm. The evolution of both the system and the environment is obtained in the weak and the strong coupling regimes, characterized respectively by damped spin oscillations and by a non-equilibrium process where the spin freezes near its initial state, which are explicitly shown to arise from a variety of reactive environmental quantum dynamics. We also explore the rich phenomenology of the intermediate coupling case, a nonperturbative regime where the system shows a complex dynamical behaviour, combining features of both the weakly and the strongly coupled case in a sequential, time-retarded fashion. Our work demonstrates the potential of TDVMPS methods for exploring otherwise elusive, nonperturbative regimes of complex open quantum systems, and points to the possibilities of exploiting the qualitative, real-time modification of quantum properties induced by non-equilibrium bath dynamics in ultrafast transient processes.  
\end{abstract}

\maketitle

\section{Introduction}

The dynamics of open quantum systems have recently attracted considerable attention in a wide range of physical, chemical and biological nanostructures where ultrafast experimental probes are increasingly able to access the real-time \emph{emergence} of processes such as energy relaxation, decoherence and dephasing, on time and length scales where standard Markovian master equations are no longer adequate \cite{Weiss1999quantum,BreuerPetruccione}. Understanding the transient, non-Markovian dynamics that can appear under these conditions is a significant theoretical challenge, particularly as strong coupling and non-equilibrium co-evolution of the many-body \emph{environment} may lead the system through qualitatively different regimes of effective behaviour \emph{over} the duration of the dynamics \cite{de2017dynamics}. However, exploiting the multi-scale complexity of such dissipative dynamics, such as the delocalisation-to-localisation transition and/or time-dependent decoherence and relaxation rates, might offer effective routes to efficient new technologies, as has been recently suggested in organic photovoltaic (OPV) materials \cite{GelinasScience2014,YaoPRB2015, bredas2017photovoltaic}, photosynthetic light harvesting proteins \cite{Fassioli20130901,chin2013role,fidler2014dynamic,romero2017quantum,scholes2017using}, quantum metrology \cite{ChinPRL2012} and various quantum information platforms and protocols \cite{XiongSciRep2015}. 

Attempts to simulate and understand the complex, non-perturbative dynamical regimes are usually carried out by means of simple benchmark models, one of the most popular being the spin-boson model (SBM) \cite{LeggettRMP1987,VojtaPhilMag2006}. Despite its apparent simplicity, the dynamics of this model remains to be fully understood, especially in the case of a sub-Ohmic spectral density where the \emph{slowly evolving} environment modes may modify the spin dynamics over the course of the relaxation \cite{LeggettRMP1987,Weiss1999quantum}, which is highly relevant for the examples described above.  As a consequence, many recent works have addressed the sub-ohmic SBM dynamics by means of advanced numerical techniques, such as hierarchical equations of motion \cite{DuanPRB2017}, time-dependent numerical and density matrix renormalization group \cite{AndersPRL2007, prior2010efficient}, path integral \cite{nalbach2010ultraslow,kast2013persistence}, Multi-Configuration Time-Dependent Hartree method \cite{ WangChemPhys2010} and various other approaches \cite{LeggettRMP1987,Weiss1999quantum,YaoPRE2013,wang2016variational,sun2016variational}. In the last years, another powerful technique has emerged that exploits the powerful physical insights and optimised codes based on the so-called Matrix Product States, which have driven numerous advances in the general simulation of cutting edge problems in quantum many-body theory and have very recently begun to be applied to microscopic open system models \cite{VerstraetePRL2004,VerstraetePRL2010, SchroderPRB2016, wall2016simulating}. In this article, we employ the recently developed Time-Dependent Variational MPS (TDVMPS) algorithm for open quantum systems \cite{haegeman2016unifying,SchroderPRB2016} to explore the different dynamical regimes of the deep sub-ohmic SBM by an explicit simulation of the complete spin-environment wavefunction. By calculating the evolution of \emph{both} the system and the environment, we uncover previously unobserved spin dynamical regimes, which we show to arise from strong back-action of various types of slow, non-equilibrium quantum dynamics in the environment that can be qualitatively tuned by spin parameters and/or spin-environment coupling strength.  

\section{Model and implementation}

\subsection{Sub-Ohmic spin-boson model}

The central component in the SBM is a single spin $1/2$ described as a two-level system. Such spin is surrounded by an environment composed by a continuum of independent harmonic oscillators whose physical origin, i.e. phonons, photons, molecular vibrations, etc., need not be specified. The SBM Hamiltonian is given by ($\hbar = 1$)
\begin{equation}\label{H0}
\begin{split}
& H = H_s + H_B+H_\text{int} = \\&\hspace{-0.07cm}-\frac{\Delta}{2}\sigma_x +\hspace{-0.07cm}\frac{\epsilon}{2}\sigma_z \hspace{-0.05cm}+\hspace{-0.1cm} \sum_n\hspace{-0.07cm}\omega_nb^\dagger_nb_n +\hspace{-0.05cm} \frac{\sigma_z}{2}\hspace{-0.05cm}\sum_ng_n\hspace{-0.05cm}\left(b^\dagger_n + b_n\right)\hspace{-0.05cm}.
\end{split}
\end{equation}
The first two terms above correspond to the Hamiltonian $H_s$ of the bare spin, the operators $\sigma_i$ being the usual Pauli matrices. Here, the energy difference between the two levels and the coherent tunnelling between them are quantified, respectively, by the bias $\epsilon$ and the rate $\Delta$. The third term, $H_B$, describes the energy of the independent harmonic oscillators through the bosonic creation and annihilation operators $b_n^\dagger$ and $b_n$. Finally, the last term in Eq. \ref{H0} contains the spin-bath interaction, where $\sigma_z$ is linearly coupled to the displacement of mode $n$ with coupling strength $g_n$.

The system-bath interaction is globally characterized by the spectral density $J(\omega) = \pi\sum_ng_n^2\delta(\omega-\omega_n)$ for which we will choose the commonly used power law  $J(\omega) = 2\pi\alpha\omega_c^{1-s}\omega^s \theta(\omega-\omega_c)$, where the hard cutoff at $\omega =\omega_c$ allows us to reproduce also the effect of finite bandwidth environments. The exponent $s$ of $J(\omega)$ is extremely important, as it determines the relative importance of high and low frequency environmental modes. For the sub-Ohmic case ($s<1$) the predominance of the slow modes induces a more complex behaviour, such as a magnetisation (localisation) quantum phase transition in the bias-free ($\epsilon=0$) SBM ground state $(\langle\sigma_z\rangle_\text{GS}\neq 0)$, which has been investigated analytically and with advanced numerics\cite{blunden2017anatomy,BullaPRL2003,ChinPRL2011,VojtaPhilMag2006,WeichselbaumPRB2009,WinterPRL2009,YaoPRE2013}. Aside from the parameter $s$, the dimensionless constant $\alpha$ determines the absolute system-bath coupling strength, which can also be expressed through the bath reorganization energy scale, $\lambda = (4\pi)^{-1}\int d\omega J(\omega)/\omega = \alpha\omega_c/2s$. 

The study of the SBM is particularly challenging for deep sub-Ohmic spectral densities, since usual perturbative approaches are bound to fail; the slow `soft' modes can become strongly displaced by the spin coupling (non-perturbative) and yet also relax slowly during the spin dynamics (non-Markovian). \cite{LeggettRMP1987,NalbachPRB2010}. Indeed, it was previously thought that large-displacement polaronic dressing of the spin tunnelling would localise (magnetise) the spin for any finite coupling at $\epsilon=0$ \cite{LeggettRMP1987}, but coherent dynamics have been shown to exist in many later studies \cite{nalbach2010ultraslow,kast2013persistence,DuanPRB2017, YaoPRE2013,SchroderPRB2016,AndersPRL2007,WangChemPhys2010,wang2016variational,sun2016variational}. Recently, it was also shown that sub-Ohmic environments both renormalise tunnelling \emph{and} develop a net displacement aligned with the spontaneous magnetisation above the ground state transition ($\alpha>\alpha_{c}$), with the latter arising from slow modes below a new interaction-dependent energy scale \cite{ChinPRL2011, blunden2017anatomy}. Additionally, Nalbach and Thorwart demonstrated with path integrals that an initially displaced sub-Ohmic environment acts as a slowly evolving energy level bias on an otherwise unbiased spin, modifying its dynamics while the bath comes to equilibrium \cite{nalbach2010ultraslow}, although no explicit bath observables were computed. Here, in order to explore the \emph{dynamical} impact of these slowly evolving modes explicitly, we explore the SBM at finite bias ($\epsilon\neq0$), as this is the situation typically found for directed energy and charge funnelling in effective light-harvesting systems, and allows the slow modes to develop bias-altering net displacements \emph{during} the course of the spin relaxation. To further emphasise these dynamics, we shall focus on the deep sub-ohmic regime, $s=0.1$. 

\begin{figure*} 
	\center
	\includegraphics[width=\linewidth]{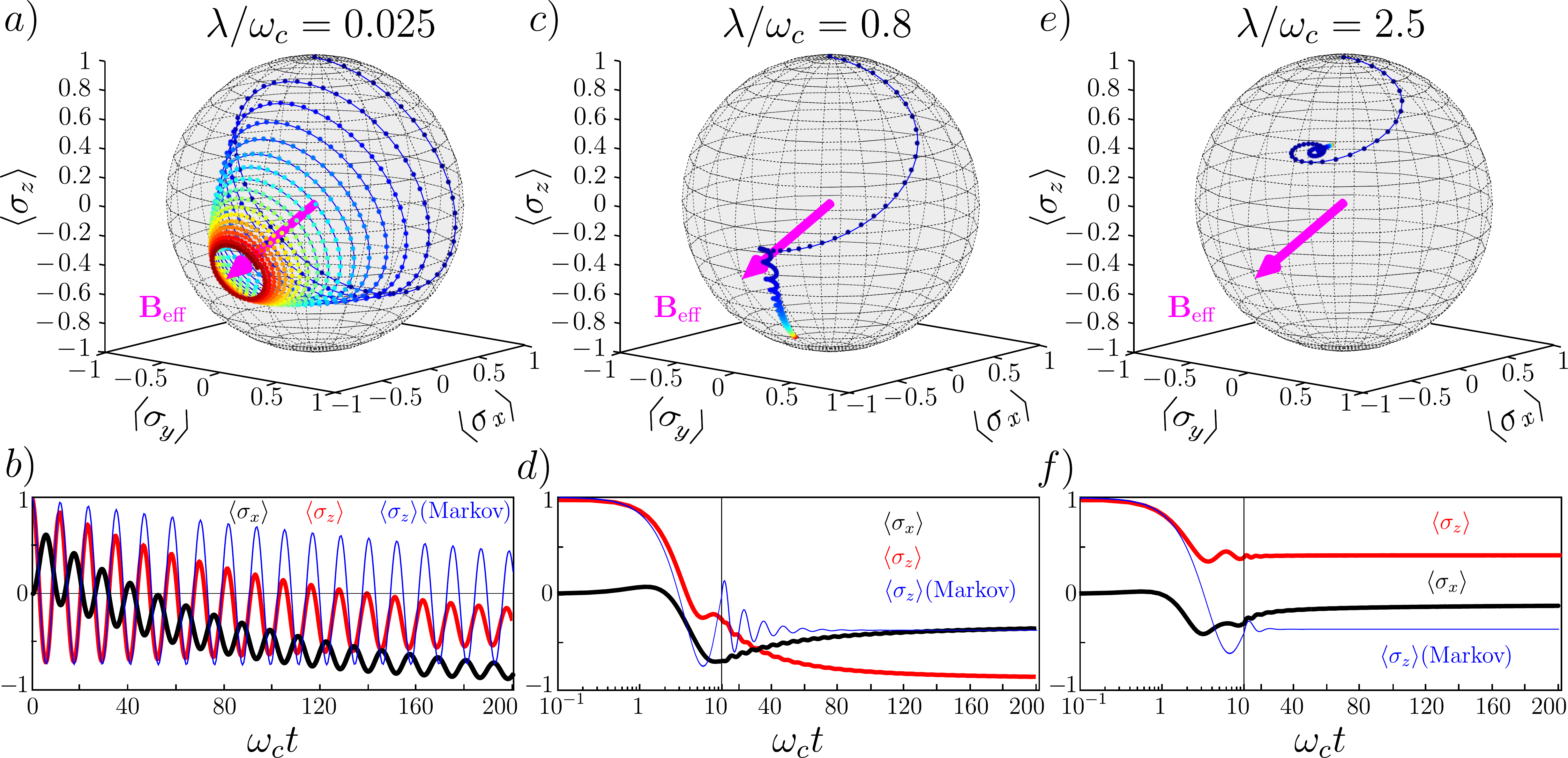}
	\vspace{-0.2cm}
	\caption{(Colour online) Time evolution of the spin on the Bloch sphere (a,c,e) and of the mean values $\langle \sigma_z\rangle$ and $\langle \sigma_x\rangle$ (b,d,f), for different system-bath reorganization energies $\lambda$ and a deep sub-ohmic spectral density, $s=0.1$. The regimes of weak ($\lambda/\omega_c = 0.025$), intermediate ($\lambda/\omega_c = 0.8$), and strong coupling ($\lambda/\omega_c = 2.5$) are shown in panels a-b, c-d, and e-f, respectively. The pink arrows in each of the Bloch spheres show the direction of the effective field around which an uncoupled spin would precess, $\mathbf{B}_\text{eff}$. In panels (b,d,f), the red and blue curves show the magnetization  as obtained with the TDVMPS algorithm and a Markovian Master Equation, respectively.}
	\label{fig1}
\end{figure*}

\subsection{Chain representation and TDVMPS parameters}

Since the TDVMPS approach is optimal for one dimensional systems \cite{VerstraetePRL2004}, we carry out a chain transformation, which maps the original bosonic environment into a one-dimensional semi-infinite chain. 
We will briefly outline the procedure here, addressing the reader to Refs. \cite{PriorPRL2010,ChinJMP2010,WoodsJMP2014} for a detailed deduction of such transformation. The first step is to conveniently choose a set of orthogonal polynomials, based on which we can rewrite the SBM Hamiltonian Eq. \ref{H0} as
\begin{equation}\label{H1}
\begin{split}
H &= H_s + \frac{\sigma_z}{2}c_0\left(a_0 + a_0^\dagger\right) + \\
&+ \sum_{k=0}^{\infty}\omega_ka^\dagger_ka_k+t_k\left(a_k^\dagger a_{k+1}+ H.c.\right),
\end{split}
\end{equation}
where the new set of bosonic operators, $a_k$, is constructed from the original ones, $b_n$, in a recursive fashion \cite{SchroderPRB2016}. Conveniently, in this representation the spin is coupled only to the first site of the chain. Moreover, for the chosen spectral density, the tight-binding coefficients $c_0$ and $\lbrace \omega_k,t_k\rbrace$ can be calculated analytically, allowing us to determine the system evolution for Hamiltonian \ref{H1} and, afterwards, apply the inverse chain mapping to obtain the dynamics of the original modes $b_k$.

The one-dimensional structure of the chain Hamiltonian Eq. \ref{H1} allows us to solve the dynamics of \textit{both} the spin and the environment by means of the TDVMPS algorithm for open systems developed in Ref. \cite{SchroderPRB2016}, a technique never  applied before to the biased sub-Ohmic SBM. It is necessary for this purpose to truncate the infinite bosonic chain at a site $k_\text{max}=L$. Note that this truncation does not introduce a systematic error in the dynamics of the whole system, but only an upper cutoff in simulation time. Indeed, the algorithm is numerically exact for times $t< \tau \propto L/\omega_c$, above which the propagating phonons are reflected backwards at the end of the chain, the resulting interference introducing artefacts in the inverse chain mapping. In the case of the spin dynamics, these errors arise only after the reflected component travels back to the site $k=0$, therefore being exact for twice as long, $t<2\tau$. Aside from the chain truncation, we use an optimized method for dynamically truncating the infinite basis of each bosonic chain site (the so-called Optimized Boson Basis), based on the early work by Guo et al. \cite{guo2012critical}. 
Ref. \cite{SchroderPRB2016} demonstrates the fast convergence and numerical accuracy of this TDVMPS approach for the unbiased SBM dynamics, including the critical regimes found for Ohmic and sub-Ohmic environments, which were found to reproduce all previous known numerical results across the entire parameter space. For the results displayed here, we obtained convergence for a chain of $L=120$ sites, MPS bond dimension $D_\text{max} = 10$, and local basis of dimension $d_\text{max} = 32$ for each chain site.

\section{System dynamics}

\subsection{Spin evolution}

The calculated spin evolution is displayed in Fig. \ref{fig1} for an initial state $\vert \!\uparrow\rangle\otimes\vert \emptyset\rangle$, where $ \vert \emptyset\rangle$ is the phononic vacuum. As a proof of principle we choose the spin parameters as $\Delta = 0.5$ and $\epsilon = 0.2$, although a further exploration of the parameter space can be found in section \ref{secdiffdyn}. First, we show the weak coupling situation, $\lambda/\omega_c =0.025$, in panels a and b. The evolution of the spin shows some of the characteristics one would expect from a Markovian Master Equation, namely a damped precession around the effective magnetic field $\mathbf{B}_\textbf{eff} \propto -\Delta \mathbf{u}_x-\epsilon\mathbf{u}_z$, with a frequency $\Omega = \sqrt{\Delta^2+\epsilon^2}/2 \approx 0.55\omega_c$ and a damping rate proportional to $\lambda$ \cite{Weiss1999quantum}. Despite these similarities, it is clear that the evolution is only well approximated by a Markov approximation for very short times (see panel \ref{fig1}b). Indeed, as time increases, two different deviations can be observed in the magnetization $\langle\sigma_z\rangle$. First, a slight reduction on the oscillation frequency, associated with the usual renormalization of the tunnelling energy $\Delta$ \cite{LeggettRMP1987}. Second, an asymmetric damping of the oscillations (the results `agree' for longer in the negative spin direction), which shifts the asymptotic value $\langle \sigma_z\rangle (t\to \infty)\equiv \langle \sigma_z\rangle_\infty$ slightly below the Markovian prediction. As we will certify below, such a damping stems from the development of a weak additional bias field along the $z$ axis, induced by large dynamical net displacements of the low-frequency bosonic environment that appear during the spin-relaxation. This would be absent in the commonly studied zero-bias case ($\epsilon = 0$), already showing how the deeply sub-ohmic, biased SBM is pathological regarding the Markov approximation, which fails to describe the dynamics even in the weak coupling case due to this unavoidable environmental bias renormalisation.

\begin{figure*}
	\center
		\includegraphics[width=\linewidth]{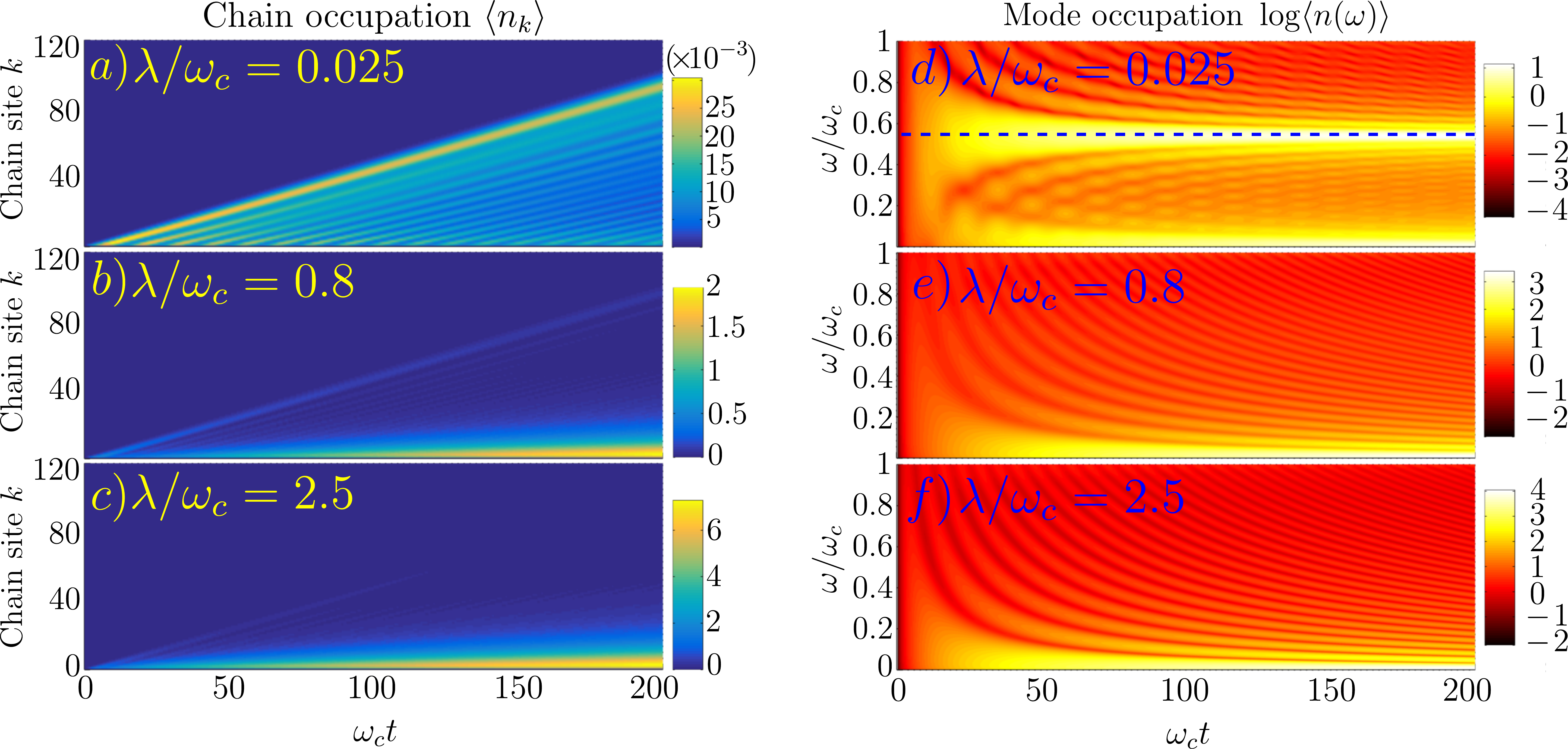}
	\caption{(Colour online) Occupation of the environment modes in the chain (a-c) and real-space (d-f) representations, as a function of mode frequency $\omega$ and time $t$, for the same values of $\lambda$ displayed in Fig. \ref{fig1}.
		The dashed line in panel d indicates the oscillation frequency of an uncoupled spin, $\Omega \approx \sqrt{\Delta^2+\epsilon^2}/2 \approx 0.55\omega_c$. }
	\label{fig2}
\end{figure*}

The situation changes significantly when the coupling increases to $\lambda/\omega_c = 0.8$ (Fig. \ref{fig1}c-d). Here, two different dynamical regimes can be clearly observed. At short times, the spin tends to precess on the Bloch sphere in a similar fashion as in the weakly coupled case, and in fairly good agreement with the Markov approximation, indicating that the slow modes are not yet participating significantly in the dynamics. However, due to the larger coupling $\lambda$, after the slow modes become activated it becomes energetically favourable for the spin to be significantly dressed by low frequency phonons. The resulting renormalization prevents the emission of the fast phonons and rapidly damps the coherent spin dynamics which initially relax in the direction of the effective spin magnetic field. This is followed by an additional effective mean field (see below), now obviously more intense than in the weak coupling scenario, which drags the spin adiabatically towards the south pole of the Bloch Sphere, $\langle \sigma_z\rangle_\infty \approx -1$. Indeed, in this later regime the spin evolution, being entirely controlled by the bath, shows a direct imprint of the bath dynamics, namely weak oscillations at $\omega_{c}$ that can also be seen in the bath's two-time displacement correlation function, which are a consequence of the hard cut-off used for $J(\omega)$. This nonperturbative `two-regime' dynamical evolution of the biased spin displayed in Figs. \ref{fig1}c-d characterizes this intermediate coupling regime where, as opposed to other situations such as the Ohmic and super-Ohmic SBMs that can also show non-Markovian features due to renormalisation effects \cite{KennesPRL2013,StrathearnARXIV2017}, the non-Markovian behaviour arises from the slow development of large net bath displacements \cite{NalbachPRB2010}, which strongly deviate the dynamics from simple master equation descriptions.

This regime nicely illustrates the basic idea of how a properly ‘programmed’ environment response could be used to qualitatively alter quantum system dynamics over the course of a process: engineering structured spectral densities, or temporal response functions could allow controlled changes in the effective Hamiltonian, i.e. the bias in the SBM, which could, to a certain extent, mimic the application of external control fields whilst also guiding the state via on non-unitary, i.e. dissipative, dynamics.  For example, the relatively simple emergent behaviour that we have shown in the sub-Ohmic SBM, i.e. the slow growth of a spin-localising bias field after relaxation, may be advantageous for processes such as charge separation in organic photovoltaics. There, it is suggested that rapid coherent dynamics may allow charges to escape their strong Coulomb interaction, whilst subsequent localisation strongly suppresses their recapture and recombination\cite{GelinasScience2014,YaoPRB2015, bredas2017photovoltaic}. 
However the true potential for harnessing tailored environmental responses could manifest in more complex, multi-stage processes, such as electron transfer (ET) in biology. For example, photosynthetic charge separation occurs by sequential electron transfers between molecules on descending timescales between ps-$\mu$s \cite{blankenship2013molecular}. As ET is very sensitive to both electronic ("system") and molecular environmental conditions, coupled, non-equilibrium and history dependent evolution of the system and environment could be used to optimise transfers at each stage, or prevent unwanted back reactions, for instance. Intriguingly, structural fluctuations of protein nanostructures can support non-equilibrium relaxation dynamics over a huge dynamical range (fs-ms), potentially allowing certain molecular motions to become relevant on timescales matching the diversity of ET rates \cite{Hu2015}, the importance of which has been discussed for classical ET kinetics \cite{Matyushov2013}.  This dynamical range is a consequence of both the vibrational properties of the component pigments \emph{and} the crystal structure-dependent dynamics of their supramolecular scaffolding, possibly suggesting clues to how similar environmental functionality, perhaps even including the co-ordination of quantum dynamics, might be achieved or explored in man-made devices by advanced nanostructure/material fabrication techniques. However, this is likely to be a very challenging feat, both experimentally and theoretically. This has partly motivated the restriction of our study to the sub-Ohmic SBM, which might be controllably realised in artificial quantum simulators, such as those proposed by in trapped ions\cite{PorrasPRA2008} or superconducting circuits\cite{YuScienceCHINA2012}.

Finally, the strongly coupled situation, $\lambda/\omega_c = 2.5$, is characterised by a practically frozen spin, as illustrated in Figs. \ref{fig1}e-f. Contrary to the previous situations, the long time magnetization $\langle \sigma_z\rangle_\infty$ remains close to its initial value, indicating the rapid evolution of the system-bath to an essentially metastable non-equilibrium energy state that is separated from the global minimum by a very large, bath-induced barrier. The tunnelling renormalization occurs in a very short timescale due to the environment reorganisation and renormalisation of the spin Hamiltonian, the spin dynamics playing a minor role in the process of energy minimization. In this regime, the dynamics more closely resemble the Marcus theory of electron transfer, and we show that the localisation can be removed by changing the bias to create a `barrierless' energy landscape \cite{Weiss1999quantum}. On the other hand, we note that the evolution of system observables towards metastable equilibria is a characteristic trait of non-adiabatic phase transitions, and has been the focus of many recent physics studies \cite{VojtaPhilMag2006,HenrietPRB2016}. 


\subsection{Dynamics of the environment}

The dynamics of the SBM summarized in Fig. \ref{fig1} have been explained in terms of the behaviour not only of the spin but of the environmental modes. The TDVMPS method automatically offers a way to verify this, as the full environmental dynamics are also available. We thus display the occupation of the phononic bath as a function of time in Fig. \ref{fig2} for the three dynamical regimes identified above. 
The left panels show the phonon occupation in the transformed system, i.e. the chain representation where the spin occupies the site $k=0$. Such a representation is a useful intermediate step toward the study of the environmental occupation, since it allows us to observe the real-time propagation of the phononic excitations along the chain. The difference between the three dynamical regimes displayed in Fig. \ref{fig1}
can be clearly observed. First, in the weak coupling case (panel \ref{fig2}a) the dynamics is largely mediated by the spin oscillations, characterized by a periodic emission of phononic wavepackets into the chain as such oscillations relax to the (environment-shifted) effective magnetic field. When the coupling is increased (panel \ref{fig2}b), a double dynamical regime arises naturally as a competition between two processes, namely the short-time phonon emission mediated by the fast modes, and the later large dressing of the spin which strongly suppress the spin oscillations. Finally, the strong coupling case (panel \ref{fig2}c) is characterized by a strong dressing of the spin even at short times, and the consequent negligible role of the fast modes.
Note that, for all the three left panels, the chosen chain length $L = 120$ is large enough to avoid any reflections below $\omega_c t = 200$, confirming that all the results displayed in this work are numerically exact. 

\begin{figure} 
	\center
	\includegraphics[width=\linewidth]{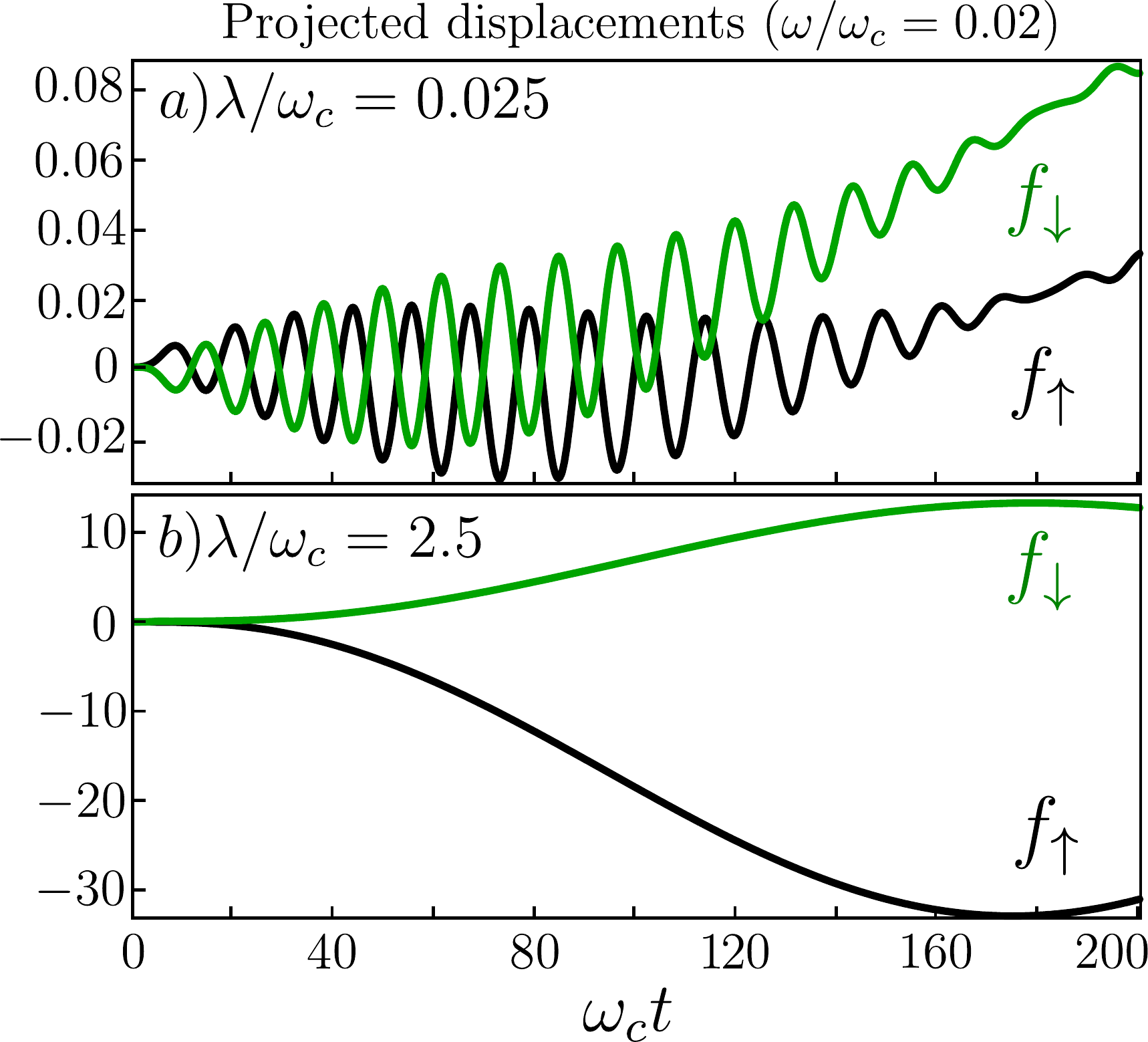}
	\vspace{-0.2cm}
	\caption{(Colour online) Spin projected displacements $f_\uparrow$ and $f_\downarrow$ of the environmental mode of frequency $\omega = 0.02\omega_c$, for the weak and the strong coupling cases (panels a and b respectively).}
	\label{fig3}
\end{figure}

\begin{figure*} 
	\center
	\includegraphics[width=0.8\linewidth]{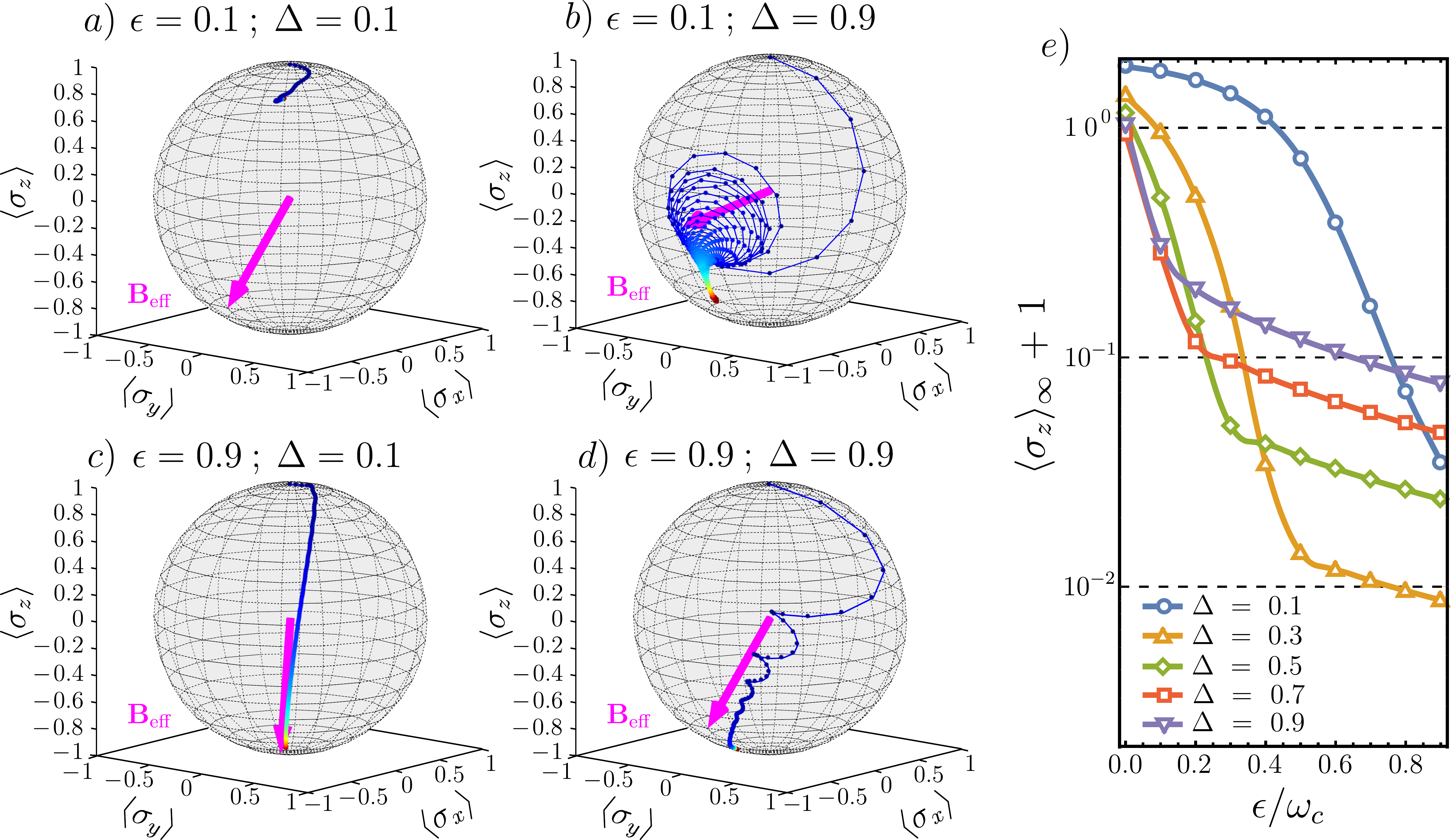}
	\vspace{-0.2cm}
	\caption{(Colour online) Time evolution of the spin on the Bloch sphere (a-d) in the intermediate coupling regime $\lambda/\omega_c = 0.8$, for different combinations of the spin parameters $\epsilon$, $\Delta$. e) Long-time magnetization $\langle \sigma_z\rangle_\infty +1$ as a function of $\Delta,\epsilon$.}
	\label{fig4}
\end{figure*}

A much better insight on the bath dynamics can be obtained by inverting the chain transformation and taking the environment into its original representation, a case we display in the right panel of Fig. \ref{fig2}.
In the weak coupling regime (panel \ref{fig2}d), we can observe a peak at energy $\Omega$, generated by the single quantum eventually emitted during the Markovian-like relaxation of the spin oscillations (direct relaxation in the eigenstate basis of $H_{S}$). The non-Markovian corrections to the spin dynamics stem from the unavoidable later population of the low-frequency modes due to the finite $\epsilon$, which can be seen in  panel \ref{fig2}d. In panels \ref{fig2}e and \ref{fig2}f, the renormalisation of the spin by the bath prevents emission of quanta at a well-defined energy (frequency), and the bath population is increasingly dominated by the slow modes and macroscopic bath reorganisation. In the chain representation used for the TDVMPS simulations, the change from emission of quanta to reorganisation-controlled dynamics appears as a switch over from propagating excitations along the chain to a strongly populated `cloud' of bosons that remains localised around the spin site. The intermediate case clearly shows both weak emission at early times, followed by the formation of localised bath excitations that arise on the same timescale as the dynamical switch-over in spin dynamics seen in Fig. \ref{fig1} b (see  Fig. \ref{fig2}b). We note the modes populations for the largest $\lambda$ exceed $10^4$, further demonstrating the need for a non-perturbative technique to explore these dynamics, and the success of TDVMPS in providing this for open systems.

In order to gain a deeper understanding of the system dynamics, we also render in Fig. \ref{fig3} the spin-projected displacements of the environment \cite{SchroderPRB2016}, defined as
\begin{equation}
f_{\uparrow,\downarrow}(n) = \left\langle\frac{ \mathbb{1}\pm \sigma_z}{2}\frac{b^\dagger_n+b_n}{2}\right\rangle,
\end{equation} 
for a representative `slow' mode at frequency $\omega_n = 0.02\omega_c$. Note that according to the above definition, the net (spin-independent) displacement of the bosonic mode $n$ is given by $f_\uparrow(n)+f_\downarrow(n)$. The behaviour of the projected displacements in the weak coupling case is depicted in panel \ref{fig3}a. Here, the anti-correlated oscillations at short times are a consequence of the projection operation onto the spin population and average out to zero, indicating no net displacement of the environment or, equivalently, the very small role played by the slow phonons. As time increases, however, these modes experience a net displacement toward the state $\vert \!\downarrow\rangle$, as the environment tries to minimize the energy of the final spin state that is already close to its Markovian long-time limit $\propto \mathbf{B}_\mathbf{eff}$. The displacement of the bosonic cloud generates an extra contribution to such effective field, tipping the spin vector slightly towards the south pole and inducing the non-Markovian deviations discussed in Fig. \ref{fig1}. This mean field picture is confirmed by the spin-insensitive character of the long-time upwards drift in $f_{\uparrow,\downarrow}$. Interestingly, polaronic correlations, see below, would show anti-correlated spin-projected displacements, and there is some anticorrelation at intermediate times ($\omega_{c}t\approx80$), indicating a \emph{dynamical change} of the bath behaviour in reaction to the spin dynamics. We can thus confirm the \emph{non-Markovian} weak coupling regime as having a slowly growing, environment-induced, effective bias field in the $z$ direction which slightly deviates the effective spin magnetic field from its original direction.

The dynamics of the projected displacements are drastically different in the strong coupling case (panel \ref{fig3}b). In this situation, the projected displacements of the bath appear at short times, as even the slow modes become rapidly displaced, showing strong polaronic (anticorrelated) displacements which are equal in magnitude, so no net displacement appears (the differences in panel \ref{fig3}b are due to weighting by the unequal spin populations). This suppresses spin tunnelling and, even for a non-zero bias $\epsilon$, the spin remains frozen in a local metastable minimum where zero-point fluctuations of the environment are unable to traverse the energetic barrier to the more stable state formed by the dressed spin down. At finite temperatures, classical fluctuations should lead to an activated relaxation with a bias-dependent activation energy\cite{Weiss1999quantum,LeggettRMP1987}.  

\subsection{Sensitivity of the intermediate coupling regime to the spin parameters}\label{secdiffdyn}

A final discussion is in place regarding the sensitivity of the intermediate-coupling regime to changes in the spin parameters $\epsilon$ and $\Delta$.
As we have discussed above, this regime arises from the competition between two processes, namely the environment-induced renormalization given by the large $\lambda$, and the damped spin oscillations at a frequency $\Omega = \sqrt{\epsilon^2+\Delta^2}/2$. If such picture is correct, one would expect the double dynamical regime to only arise when the inverse timescales $\lambda^{-1}$ and $\Omega^{-1}$ are comparable. As a consequence, for a fixed $\lambda = 0.8$ (corresponding to the intermediate coupling regime for $\epsilon = 0.2$, $\Delta = 0.5$), the dynamics are expected to be extremely sensitive to the particular values of $\epsilon$ and $\Delta$. This sensitivity is certified in Fig. \ref{fig4}, which clearly evidences the interplay between the two aforementioned mechanisms. Indeed, note that the spin dynamical evolution becomes increasingly similar to that of the strong and the weak coupling regimes, in the limits of larger (panel a) and smaller (panel d) ratios $\lambda/\Omega$ respectively.

Aside from the aforementioned change of regime as $\Omega$ changes, the system dynamics can also drastically change depending on the relative values of $\epsilon$ and $\Delta$. As shown in panels \ref{fig4}b-c, a large bias to tunnelling ratio $\epsilon/\Delta$ results in a reduction of the oscillations, allowing from an almost straight spin flip through the centre of the Bloch sphere. On the other hand, a small ratio $\epsilon/\Delta$ effectively delays the onset of the bath-induced renormalization. Although expected, both these results allow for an extra degree of flexibility over the spin dynamics, which could be exploited for state preparation in the Bloch sphere. A simple example could be the harnessing of this effect to perform an irreversible spin flip. In panel \ref{fig4}e we display the final magnetization as a function of $\epsilon$ and $\Delta$, showing that such a spin flip can reach fidelities larger than $99\%$. This is just one of the potential applications of environment engineering, a research area to which, as we have shown, optimised numerical algorithms such as TDVMPS can contribute significantly.

\section{Conclusion}

The results above, spanning the entire weak-to-strong phenomenology of the biased deep sub-Ohmic regime, demonstrate how the evolving state of the bath determines the fate of the spin in a way that self-consistently depends on the history of the spin dynamics, which we have observed directly through the environment observables. This occurs due to the comparable timescales of bath and system evolution, which permit, especially in the intermediate regime, a period of nearly free, coherent spin dynamics that allow exploration of the Bloch Sphere before the strong environmental effects take over the subsequent dynamics. However, the bath-induced dynamics depend on the prior spin motion and can consequently be tuned by the bare spin parameters of $H_{S}$, particularly the bias that can induce a reinforcing environmental field, even at very weak coupling. Having further demonstrated a powerful technique capable of handling the non-perturbative nature of these physics, we believe that a fruitful avenue of future investigation would be to understand how the complex environmental `feedback' dynamics that modify coherent quantum processes could be controlled and exploited in engineered solid-state nanostructures \cite{scholes2017using}, either passively, i.e. through bath engineering, or actively through quantum control protocols \cite{eckel2009coherent}.

\begin{acknowledgements}
  CGB acknowledges the Spanish MECD (FPU13/01225 fellowship) and the support of F. J. Garcia Vidal. AWC and FAYNS acknowledge the Winton Programme for the Physics of Sustainability for support.
\end{acknowledgements}

\bibliographystyle{apsrev4-1}
\bibliography{BIBLIO}

\begin{thebibliography}{48}%
\makeatletter
\providecommand \@ifxundefined [1]{%
 \@ifx{#1\undefined}
}%
\providecommand \@ifnum [1]{%
 \ifnum #1\expandafter \@firstoftwo
 \else \expandafter \@secondoftwo
 \fi
}%
\providecommand \@ifx [1]{%
 \ifx #1\expandafter \@firstoftwo
 \else \expandafter \@secondoftwo
 \fi
}%
\providecommand \natexlab [1]{#1}%
\providecommand \enquote  [1]{``#1''}%
\providecommand \bibnamefont  [1]{#1}%
\providecommand \bibfnamefont [1]{#1}%
\providecommand \citenamefont [1]{#1}%
\providecommand \href@noop [0]{\@secondoftwo}%
\providecommand \href [0]{\begingroup \@sanitize@url \@href}%
\providecommand \@href[1]{\@@startlink{#1}\@@href}%
\providecommand \@@href[1]{\endgroup#1\@@endlink}%
\providecommand \@sanitize@url [0]{\catcode `\\12\catcode `\$12\catcode
  `\&12\catcode `\#12\catcode `\^12\catcode `\_12\catcode `\%12\relax}%
\providecommand \@@startlink[1]{}%
\providecommand \@@endlink[0]{}%
\providecommand \url  [0]{\begingroup\@sanitize@url \@url }%
\providecommand \@url [1]{\endgroup\@href {#1}{\urlprefix }}%
\providecommand \urlprefix  [0]{URL }%
\providecommand \Eprint [0]{\href }%
\providecommand \doibase [0]{http://dx.doi.org/}%
\providecommand \selectlanguage [0]{\@gobble}%
\providecommand \bibinfo  [0]{\@secondoftwo}%
\providecommand \bibfield  [0]{\@secondoftwo}%
\providecommand \translation [1]{[#1]}%
\providecommand \BibitemOpen [0]{}%
\providecommand \bibitemStop [0]{}%
\providecommand \bibitemNoStop [0]{.\EOS\space}%
\providecommand \EOS [0]{\spacefactor3000\relax}%
\providecommand \BibitemShut  [1]{\csname bibitem#1\endcsname}%
\let\auto@bib@innerbib\@empty
\bibitem [{\citenamefont {Weiss}(1999)}]{Weiss1999quantum}%
  \BibitemOpen
  \bibfield  {author} {\bibinfo {author} {\bibfnamefont {U.}~\bibnamefont
  {Weiss}},\ }\href@noop {} {\emph {\bibinfo {title} {Quantum dissipative
  systems}}},\ Vol.~\bibinfo {volume} {10}\ (\bibinfo  {publisher} {World
  Scientific},\ \bibinfo {year} {1999})\BibitemShut {NoStop}%
\bibitem [{\citenamefont {Breuer}\ and\ \citenamefont
  {Petruccione}(2002)}]{BreuerPetruccione}%
  \BibitemOpen
  \bibfield  {author} {\bibinfo {author} {\bibfnamefont {H.}~\bibnamefont
  {Breuer}}\ and\ \bibinfo {author} {\bibfnamefont {F.}~\bibnamefont
  {Petruccione}},\ }\href@noop {} {\emph {\bibinfo {title} {The Theory of Open
  Quantum Systems}}}\ (\bibinfo  {publisher} {Oxford University Press},\
  \bibinfo {year} {2002})\BibitemShut {NoStop}%
\bibitem [{\citenamefont {de~Vega}\ and\ \citenamefont
  {Alonso}(2017)}]{de2017dynamics}%
  \BibitemOpen
  \bibfield  {author} {\bibinfo {author} {\bibfnamefont {I.}~\bibnamefont
  {de~Vega}}\ and\ \bibinfo {author} {\bibfnamefont {D.}~\bibnamefont
  {Alonso}},\ }\href
  {https://journals.aps.org/rmp/abstract/10.1103/RevModPhys.89.015001}
  {\bibfield  {journal} {\bibinfo  {journal} {Reviews of Modern Physics}\
  }\textbf {\bibinfo {volume} {89}},\ \bibinfo {pages} {015001} (\bibinfo
  {year} {2017})}\BibitemShut {NoStop}%
\bibitem [{\citenamefont {G{\'e}linas}\ \emph {et~al.}(2014)\citenamefont
  {G{\'e}linas}, \citenamefont {Rao}, \citenamefont {Kumar}, \citenamefont
  {Smith}, \citenamefont {Chin}, \citenamefont {Clark}, \citenamefont {van~der
  Poll}, \citenamefont {Bazan},\ and\ \citenamefont
  {Friend}}]{GelinasScience2014}%
  \BibitemOpen
  \bibfield  {author} {\bibinfo {author} {\bibfnamefont {S.}~\bibnamefont
  {G{\'e}linas}}, \bibinfo {author} {\bibfnamefont {A.}~\bibnamefont {Rao}},
  \bibinfo {author} {\bibfnamefont {A.}~\bibnamefont {Kumar}}, \bibinfo
  {author} {\bibfnamefont {S.~L.}\ \bibnamefont {Smith}}, \bibinfo {author}
  {\bibfnamefont {A.~W.}\ \bibnamefont {Chin}}, \bibinfo {author}
  {\bibfnamefont {J.}~\bibnamefont {Clark}}, \bibinfo {author} {\bibfnamefont
  {T.~S.}\ \bibnamefont {van~der Poll}}, \bibinfo {author} {\bibfnamefont
  {G.~C.}\ \bibnamefont {Bazan}}, \ and\ \bibinfo {author} {\bibfnamefont
  {R.~H.}\ \bibnamefont {Friend}},\ }\href {\doibase 10.1126/science.1246249}
  {\bibfield  {journal} {\bibinfo  {journal} {Science}\ }\textbf {\bibinfo
  {volume} {343}},\ \bibinfo {pages} {512} (\bibinfo {year}
  {2014})}\BibitemShut {NoStop}%
\bibitem [{\citenamefont {Yao}(2015)}]{YaoPRB2015}%
  \BibitemOpen
  \bibfield  {author} {\bibinfo {author} {\bibfnamefont {Y.}~\bibnamefont
  {Yao}},\ }\href {\doibase 10.1103/PhysRevB.91.045421} {\bibfield  {journal}
  {\bibinfo  {journal} {Phys. Rev. B}\ }\textbf {\bibinfo {volume} {91}},\
  \bibinfo {pages} {045421} (\bibinfo {year} {2015})}\BibitemShut {NoStop}%
\bibitem [{\citenamefont {Br{\'e}das}\ \emph {et~al.}(2017)\citenamefont
  {Br{\'e}das}, \citenamefont {Sargent},\ and\ \citenamefont
  {Scholes}}]{bredas2017photovoltaic}%
  \BibitemOpen
  \bibfield  {author} {\bibinfo {author} {\bibfnamefont {J.-L.}\ \bibnamefont
  {Br{\'e}das}}, \bibinfo {author} {\bibfnamefont {E.~H.}\ \bibnamefont
  {Sargent}}, \ and\ \bibinfo {author} {\bibfnamefont {G.~D.}\ \bibnamefont
  {Scholes}},\ }\href
  {http://www.nature.com/nmat/journal/v16/n1/full/nmat4767.html} {\bibfield
  {journal} {\bibinfo  {journal} {Nature materials}\ }\textbf {\bibinfo
  {volume} {16}},\ \bibinfo {pages} {35} (\bibinfo {year} {2017})}\BibitemShut
  {NoStop}%
\bibitem [{\citenamefont {Fassioli}\ \emph {et~al.}(2013)\citenamefont
  {Fassioli}, \citenamefont {Dinshaw}, \citenamefont {Arpin},\ and\
  \citenamefont {Scholes}}]{Fassioli20130901}%
  \BibitemOpen
  \bibfield  {author} {\bibinfo {author} {\bibfnamefont {F.}~\bibnamefont
  {Fassioli}}, \bibinfo {author} {\bibfnamefont {R.}~\bibnamefont {Dinshaw}},
  \bibinfo {author} {\bibfnamefont {P.~C.}\ \bibnamefont {Arpin}}, \ and\
  \bibinfo {author} {\bibfnamefont {G.~D.}\ \bibnamefont {Scholes}},\ }\href
  {http://rsif.royalsocietypublishing.org/content/11/92/20130901} {\bibfield
  {journal} {\bibinfo  {journal} {Journal of The Royal Society Interface}\
  }\textbf {\bibinfo {volume} {11}} (\bibinfo {year} {2013})}\BibitemShut
  {NoStop}%
\bibitem [{\citenamefont {Chin}\ \emph {et~al.}(2013)\citenamefont {Chin},
  \citenamefont {Prior}, \citenamefont {Rosenbach}, \citenamefont
  {Caycedo-Soler}, \citenamefont {Huelga},\ and\ \citenamefont
  {Plenio}}]{chin2013role}%
  \BibitemOpen
  \bibfield  {author} {\bibinfo {author} {\bibfnamefont {A.}~\bibnamefont
  {Chin}}, \bibinfo {author} {\bibfnamefont {J.}~\bibnamefont {Prior}},
  \bibinfo {author} {\bibfnamefont {R.}~\bibnamefont {Rosenbach}}, \bibinfo
  {author} {\bibfnamefont {F.}~\bibnamefont {Caycedo-Soler}}, \bibinfo {author}
  {\bibfnamefont {S.}~\bibnamefont {Huelga}}, \ and\ \bibinfo {author}
  {\bibfnamefont {M.}~\bibnamefont {Plenio}},\ }\href
  {http://www.nature.com/nphys/journal/v9/n2/full/nphys2515.html} {\bibfield
  {journal} {\bibinfo  {journal} {Nature Physics}\ }\textbf {\bibinfo {volume}
  {9}},\ \bibinfo {pages} {113} (\bibinfo {year} {2013})}\BibitemShut {NoStop}%
\bibitem [{\citenamefont {Fidler}\ \emph {et~al.}(2014)\citenamefont {Fidler},
  \citenamefont {Singh}, \citenamefont {Long}, \citenamefont {Dahlberg},\ and\
  \citenamefont {Engel}}]{fidler2014dynamic}%
  \BibitemOpen
  \bibfield  {author} {\bibinfo {author} {\bibfnamefont {A.~F.}\ \bibnamefont
  {Fidler}}, \bibinfo {author} {\bibfnamefont {V.~P.}\ \bibnamefont {Singh}},
  \bibinfo {author} {\bibfnamefont {P.~D.}\ \bibnamefont {Long}}, \bibinfo
  {author} {\bibfnamefont {P.~D.}\ \bibnamefont {Dahlberg}}, \ and\ \bibinfo
  {author} {\bibfnamefont {G.~S.}\ \bibnamefont {Engel}},\ }\href
  {https://www.nature.com/articles/ncomms4286} {\bibfield  {journal} {\bibinfo
  {journal} {Nature communications}\ }\textbf {\bibinfo {volume} {5}},\
  \bibinfo {pages} {3286} (\bibinfo {year} {2014})}\BibitemShut {NoStop}%
\bibitem [{\citenamefont {Romero}\ \emph {et~al.}(2017)\citenamefont {Romero},
  \citenamefont {Novoderezhkin},\ and\ \citenamefont {van
  Grondelle}}]{romero2017quantum}%
  \BibitemOpen
  \bibfield  {author} {\bibinfo {author} {\bibfnamefont {E.}~\bibnamefont
  {Romero}}, \bibinfo {author} {\bibfnamefont {V.~I.}\ \bibnamefont
  {Novoderezhkin}}, \ and\ \bibinfo {author} {\bibfnamefont {R.}~\bibnamefont
  {van Grondelle}},\ }\href
  {https://www.nature.com/nature/journal/v543/n7645/full/nature22012.html}
  {\bibfield  {journal} {\bibinfo  {journal} {Nature}\ }\textbf {\bibinfo
  {volume} {543}},\ \bibinfo {pages} {355} (\bibinfo {year}
  {2017})}\BibitemShut {NoStop}%
\bibitem [{\citenamefont {Scholes}\ \emph {et~al.}(2017)\citenamefont
  {Scholes}, \citenamefont {Fleming}, \citenamefont {Chen}, \citenamefont
  {Aspuru-Guzik}, \citenamefont {Buchleitner}, \citenamefont {Coker},
  \citenamefont {Engel}, \citenamefont {van Grondelle}, \citenamefont
  {Ishizaki}, \citenamefont {Jonas} \emph {et~al.}}]{scholes2017using}%
  \BibitemOpen
  \bibfield  {author} {\bibinfo {author} {\bibfnamefont {G.~D.}\ \bibnamefont
  {Scholes}}, \bibinfo {author} {\bibfnamefont {G.~R.}\ \bibnamefont
  {Fleming}}, \bibinfo {author} {\bibfnamefont {L.~X.}\ \bibnamefont {Chen}},
  \bibinfo {author} {\bibfnamefont {A.}~\bibnamefont {Aspuru-Guzik}}, \bibinfo
  {author} {\bibfnamefont {A.}~\bibnamefont {Buchleitner}}, \bibinfo {author}
  {\bibfnamefont {D.~F.}\ \bibnamefont {Coker}}, \bibinfo {author}
  {\bibfnamefont {G.~S.}\ \bibnamefont {Engel}}, \bibinfo {author}
  {\bibfnamefont {R.}~\bibnamefont {van Grondelle}}, \bibinfo {author}
  {\bibfnamefont {A.}~\bibnamefont {Ishizaki}}, \bibinfo {author}
  {\bibfnamefont {D.~M.}\ \bibnamefont {Jonas}},  \emph {et~al.},\ }\href
  {http://www.nature.com/nature/journal/v543/n7647/full/nature21425.html}
  {\bibfield  {journal} {\bibinfo  {journal} {Nature}\ }\textbf {\bibinfo
  {volume} {543}},\ \bibinfo {pages} {647} (\bibinfo {year}
  {2017})}\BibitemShut {NoStop}%
\bibitem [{\citenamefont {Chin}\ \emph {et~al.}(2012)\citenamefont {Chin},
  \citenamefont {Huelga},\ and\ \citenamefont {Plenio}}]{ChinPRL2012}%
  \BibitemOpen
  \bibfield  {author} {\bibinfo {author} {\bibfnamefont {A.~W.}\ \bibnamefont
  {Chin}}, \bibinfo {author} {\bibfnamefont {S.~F.}\ \bibnamefont {Huelga}}, \
  and\ \bibinfo {author} {\bibfnamefont {M.~B.}\ \bibnamefont {Plenio}},\
  }\href {\doibase 10.1103/PhysRevLett.109.233601} {\bibfield  {journal}
  {\bibinfo  {journal} {Phys. Rev. Lett.}\ }\textbf {\bibinfo {volume} {109}},\
  \bibinfo {pages} {233601} (\bibinfo {year} {2012})}\BibitemShut {NoStop}%
\bibitem [{\citenamefont {Xiong}\ \emph {et~al.}(2015)\citenamefont {Xiong},
  \citenamefont {Lo}, \citenamefont {Zhang}, \citenamefont {Feng},\ and\
  \citenamefont {Nori}}]{XiongSciRep2015}%
  \BibitemOpen
  \bibfield  {author} {\bibinfo {author} {\bibfnamefont {H.-N.}\ \bibnamefont
  {Xiong}}, \bibinfo {author} {\bibfnamefont {P.-Y.}\ \bibnamefont {Lo}},
  \bibinfo {author} {\bibfnamefont {W.-M.}\ \bibnamefont {Zhang}}, \bibinfo
  {author} {\bibfnamefont {D.~H.}\ \bibnamefont {Feng}}, \ and\ \bibinfo
  {author} {\bibfnamefont {F.}~\bibnamefont {Nori}},\ }\href
  {http://dx.doi.org/10.1038/srep13353} {\bibfield  {journal} {\bibinfo
  {journal} {Sci Rep.}\ }\textbf {\bibinfo {volume} {5}},\ \bibinfo {pages}
  {13353} (\bibinfo {year} {2015})}\BibitemShut {NoStop}%
\bibitem [{\citenamefont {Leggett}\ \emph {et~al.}(1987)\citenamefont
  {Leggett}, \citenamefont {Chakravarty}, \citenamefont {Dorsey}, \citenamefont
  {Fisher}, \citenamefont {Garg},\ and\ \citenamefont
  {Zwerger}}]{LeggettRMP1987}%
  \BibitemOpen
  \bibfield  {author} {\bibinfo {author} {\bibfnamefont {A.~J.}\ \bibnamefont
  {Leggett}}, \bibinfo {author} {\bibfnamefont {S.}~\bibnamefont
  {Chakravarty}}, \bibinfo {author} {\bibfnamefont {A.~T.}\ \bibnamefont
  {Dorsey}}, \bibinfo {author} {\bibfnamefont {M.~P.~A.}\ \bibnamefont
  {Fisher}}, \bibinfo {author} {\bibfnamefont {A.}~\bibnamefont {Garg}}, \ and\
  \bibinfo {author} {\bibfnamefont {W.}~\bibnamefont {Zwerger}},\ }\href
  {\doibase 10.1103/RevModPhys.59.1} {\bibfield  {journal} {\bibinfo  {journal}
  {Rev. Mod. Phys.}\ }\textbf {\bibinfo {volume} {59}},\ \bibinfo {pages} {1}
  (\bibinfo {year} {1987})}\BibitemShut {NoStop}%
\bibitem [{\citenamefont {Vojta}(2006)}]{VojtaPhilMag2006}%
  \BibitemOpen
  \bibfield  {author} {\bibinfo {author} {\bibfnamefont {M.}~\bibnamefont
  {Vojta}},\ }\href {\doibase 10.1080/14786430500070396} {\bibfield  {journal}
  {\bibinfo  {journal} {Philosophical Magazine}\ }\textbf {\bibinfo {volume}
  {86}},\ \bibinfo {pages} {1807} (\bibinfo {year} {2006})}\BibitemShut
  {NoStop}%
\bibitem [{\citenamefont {Duan}\ \emph {et~al.}(2017)\citenamefont {Duan},
  \citenamefont {Tang}, \citenamefont {Cao},\ and\ \citenamefont
  {Wu}}]{DuanPRB2017}%
  \BibitemOpen
  \bibfield  {author} {\bibinfo {author} {\bibfnamefont {C.}~\bibnamefont
  {Duan}}, \bibinfo {author} {\bibfnamefont {Z.}~\bibnamefont {Tang}}, \bibinfo
  {author} {\bibfnamefont {J.}~\bibnamefont {Cao}}, \ and\ \bibinfo {author}
  {\bibfnamefont {J.}~\bibnamefont {Wu}},\ }\href {\doibase
  10.1103/PhysRevB.95.214308} {\bibfield  {journal} {\bibinfo  {journal} {Phys.
  Rev. B}\ }\textbf {\bibinfo {volume} {95}},\ \bibinfo {pages} {214308}
  (\bibinfo {year} {2017})}\BibitemShut {NoStop}%
\bibitem [{\citenamefont {Anders}\ \emph {et~al.}(2007)\citenamefont {Anders},
  \citenamefont {Bulla},\ and\ \citenamefont {Vojta}}]{AndersPRL2007}%
  \BibitemOpen
  \bibfield  {author} {\bibinfo {author} {\bibfnamefont {F.~B.}\ \bibnamefont
  {Anders}}, \bibinfo {author} {\bibfnamefont {R.}~\bibnamefont {Bulla}}, \
  and\ \bibinfo {author} {\bibfnamefont {M.}~\bibnamefont {Vojta}},\ }\href
  {\doibase 10.1103/PhysRevLett.98.210402} {\bibfield  {journal} {\bibinfo
  {journal} {Phys. Rev. Lett.}\ }\textbf {\bibinfo {volume} {98}},\ \bibinfo
  {pages} {210402} (\bibinfo {year} {2007})}\BibitemShut {NoStop}%
\bibitem [{\citenamefont {Prior}\ \emph
  {et~al.}(2010{\natexlab{a}})\citenamefont {Prior}, \citenamefont {Chin},
  \citenamefont {Huelga},\ and\ \citenamefont {Plenio}}]{prior2010efficient}%
  \BibitemOpen
  \bibfield  {author} {\bibinfo {author} {\bibfnamefont {J.}~\bibnamefont
  {Prior}}, \bibinfo {author} {\bibfnamefont {A.~W.}\ \bibnamefont {Chin}},
  \bibinfo {author} {\bibfnamefont {S.~F.}\ \bibnamefont {Huelga}}, \ and\
  \bibinfo {author} {\bibfnamefont {M.~B.}\ \bibnamefont {Plenio}},\ }\href
  {\doibase 10.1103/PhysRevLett.105.050404} {\bibfield  {journal} {\bibinfo
  {journal} {Phys. Rev. Lett.}\ }\textbf {\bibinfo {volume} {105}},\ \bibinfo
  {pages} {050404} (\bibinfo {year} {2010}{\natexlab{a}})}\BibitemShut
  {NoStop}%
\bibitem [{\citenamefont {Nalbach}\ and\ \citenamefont
  {Thorwart}(2010{\natexlab{a}})}]{nalbach2010ultraslow}%
  \BibitemOpen
  \bibfield  {author} {\bibinfo {author} {\bibfnamefont {P.}~\bibnamefont
  {Nalbach}}\ and\ \bibinfo {author} {\bibfnamefont {M.}~\bibnamefont
  {Thorwart}},\ }\href
  {https://journals.aps.org/prb/abstract/10.1103/PhysRevB.81.054308} {\bibfield
   {journal} {\bibinfo  {journal} {Physical Review B}\ }\textbf {\bibinfo
  {volume} {81}},\ \bibinfo {pages} {054308} (\bibinfo {year}
  {2010}{\natexlab{a}})}\BibitemShut {NoStop}%
\bibitem [{\citenamefont {Kast}\ and\ \citenamefont
  {Ankerhold}(2013)}]{kast2013persistence}%
  \BibitemOpen
  \bibfield  {author} {\bibinfo {author} {\bibfnamefont {D.}~\bibnamefont
  {Kast}}\ and\ \bibinfo {author} {\bibfnamefont {J.}~\bibnamefont
  {Ankerhold}},\ }\href
  {https://link.aps.org/doi/10.1103/PhysRevLett.110.010402} {\bibfield
  {journal} {\bibinfo  {journal} {Physical review letters}\ }\textbf {\bibinfo
  {volume} {110}},\ \bibinfo {pages} {010402} (\bibinfo {year}
  {2013})}\BibitemShut {NoStop}%
\bibitem [{\citenamefont {Wang}\ and\ \citenamefont
  {Thoss}(2010)}]{WangChemPhys2010}%
  \BibitemOpen
  \bibfield  {author} {\bibinfo {author} {\bibfnamefont {H.}~\bibnamefont
  {Wang}}\ and\ \bibinfo {author} {\bibfnamefont {M.}~\bibnamefont {Thoss}},\
  }\href {\doibase http://dx.doi.org/10.1016/j.chemphys.2010.02.027} {\bibfield
   {journal} {\bibinfo  {journal} {Chemical Physics}\ }\textbf {\bibinfo
  {volume} {370}},\ \bibinfo {pages} {78 } (\bibinfo {year} {2010})},\ \bibinfo
  {note} {dynamics of molecular systems: From quantum to classical}\BibitemShut
  {NoStop}%
\bibitem [{\citenamefont {Yao}\ \emph {et~al.}(2013)\citenamefont {Yao},
  \citenamefont {Duan}, \citenamefont {L\"u}, \citenamefont {Wu},\ and\
  \citenamefont {Zhao}}]{YaoPRE2013}%
  \BibitemOpen
  \bibfield  {author} {\bibinfo {author} {\bibfnamefont {Y.}~\bibnamefont
  {Yao}}, \bibinfo {author} {\bibfnamefont {L.}~\bibnamefont {Duan}}, \bibinfo
  {author} {\bibfnamefont {Z.}~\bibnamefont {L\"u}}, \bibinfo {author}
  {\bibfnamefont {C.-Q.}\ \bibnamefont {Wu}}, \ and\ \bibinfo {author}
  {\bibfnamefont {Y.}~\bibnamefont {Zhao}},\ }\href {\doibase
  10.1103/PhysRevE.88.023303} {\bibfield  {journal} {\bibinfo  {journal} {Phys.
  Rev. E}\ }\textbf {\bibinfo {volume} {88}},\ \bibinfo {pages} {023303}
  (\bibinfo {year} {2013})}\BibitemShut {NoStop}%
\bibitem [{\citenamefont {Wang}\ \emph {et~al.}(2016)\citenamefont {Wang},
  \citenamefont {Chen}, \citenamefont {Zhou},\ and\ \citenamefont
  {Zhao}}]{wang2016variational}%
  \BibitemOpen
  \bibfield  {author} {\bibinfo {author} {\bibfnamefont {L.}~\bibnamefont
  {Wang}}, \bibinfo {author} {\bibfnamefont {L.}~\bibnamefont {Chen}}, \bibinfo
  {author} {\bibfnamefont {N.}~\bibnamefont {Zhou}}, \ and\ \bibinfo {author}
  {\bibfnamefont {Y.}~\bibnamefont {Zhao}},\ }\href
  {http://aip.scitation.org/doi/full/10.1063/1.4939144} {\bibfield  {journal}
  {\bibinfo  {journal} {The Journal of chemical physics}\ }\textbf {\bibinfo
  {volume} {144}},\ \bibinfo {pages} {024101} (\bibinfo {year}
  {2016})}\BibitemShut {NoStop}%
\bibitem [{\citenamefont {Sun}\ \emph {et~al.}(2016)\citenamefont {Sun},
  \citenamefont {Fujihashi}, \citenamefont {Ishizaki},\ and\ \citenamefont
  {Zhao}}]{sun2016variational}%
  \BibitemOpen
  \bibfield  {author} {\bibinfo {author} {\bibfnamefont {K.-W.}\ \bibnamefont
  {Sun}}, \bibinfo {author} {\bibfnamefont {Y.}~\bibnamefont {Fujihashi}},
  \bibinfo {author} {\bibfnamefont {A.}~\bibnamefont {Ishizaki}}, \ and\
  \bibinfo {author} {\bibfnamefont {Y.}~\bibnamefont {Zhao}},\ }\href
  {http://aip.scitation.org/doi/full/10.1063/1.4950888} {\bibfield  {journal}
  {\bibinfo  {journal} {The Journal of chemical physics}\ }\textbf {\bibinfo
  {volume} {144}},\ \bibinfo {pages} {204106} (\bibinfo {year}
  {2016})}\BibitemShut {NoStop}%
\bibitem [{\citenamefont {Verstraete}\ \emph {et~al.}(2004)\citenamefont
  {Verstraete}, \citenamefont {Garc\'{\i}a-Ripoll},\ and\ \citenamefont
  {Cirac}}]{VerstraetePRL2004}%
  \BibitemOpen
  \bibfield  {author} {\bibinfo {author} {\bibfnamefont {F.}~\bibnamefont
  {Verstraete}}, \bibinfo {author} {\bibfnamefont {J.~J.}\ \bibnamefont
  {Garc\'{\i}a-Ripoll}}, \ and\ \bibinfo {author} {\bibfnamefont {J.~I.}\
  \bibnamefont {Cirac}},\ }\href {\doibase 10.1103/PhysRevLett.93.207204}
  {\bibfield  {journal} {\bibinfo  {journal} {Phys. Rev. Lett.}\ }\textbf
  {\bibinfo {volume} {93}},\ \bibinfo {pages} {207204} (\bibinfo {year}
  {2004})}\BibitemShut {NoStop}%
\bibitem [{\citenamefont {Verstraete}\ and\ \citenamefont
  {Cirac}(2010)}]{VerstraetePRL2010}%
  \BibitemOpen
  \bibfield  {author} {\bibinfo {author} {\bibfnamefont {F.}~\bibnamefont
  {Verstraete}}\ and\ \bibinfo {author} {\bibfnamefont {J.~I.}\ \bibnamefont
  {Cirac}},\ }\href {\doibase 10.1103/PhysRevLett.104.190405} {\bibfield
  {journal} {\bibinfo  {journal} {Phys. Rev. Lett.}\ }\textbf {\bibinfo
  {volume} {104}},\ \bibinfo {pages} {190405} (\bibinfo {year}
  {2010})}\BibitemShut {NoStop}%
\bibitem [{\citenamefont {Schr\"oder}\ and\ \citenamefont
  {Chin}(2016)}]{SchroderPRB2016}%
  \BibitemOpen
  \bibfield  {author} {\bibinfo {author} {\bibfnamefont {F.~A. Y.~N.}\
  \bibnamefont {Schr\"oder}}\ and\ \bibinfo {author} {\bibfnamefont {A.~W.}\
  \bibnamefont {Chin}},\ }\href {\doibase 10.1103/PhysRevB.93.075105}
  {\bibfield  {journal} {\bibinfo  {journal} {Phys. Rev. B}\ }\textbf {\bibinfo
  {volume} {93}},\ \bibinfo {pages} {075105} (\bibinfo {year}
  {2016})}\BibitemShut {NoStop}%
\bibitem [{\citenamefont {Wall}\ \emph {et~al.}(2016)\citenamefont {Wall},
  \citenamefont {Safavi-Naini},\ and\ \citenamefont
  {Rey}}]{wall2016simulating}%
  \BibitemOpen
  \bibfield  {author} {\bibinfo {author} {\bibfnamefont {M.~L.}\ \bibnamefont
  {Wall}}, \bibinfo {author} {\bibfnamefont {A.}~\bibnamefont {Safavi-Naini}},
  \ and\ \bibinfo {author} {\bibfnamefont {A.~M.}\ \bibnamefont {Rey}},\ }\href
  {https://journals.aps.org/pra/abstract/10.1103/PhysRevA.94.053637} {\bibfield
   {journal} {\bibinfo  {journal} {Physical Review A}\ }\textbf {\bibinfo
  {volume} {94}},\ \bibinfo {pages} {053637} (\bibinfo {year}
  {2016})}\BibitemShut {NoStop}%
\bibitem [{\citenamefont {Haegeman}\ \emph {et~al.}(2016)\citenamefont
  {Haegeman}, \citenamefont {Lubich}, \citenamefont {Oseledets}, \citenamefont
  {Vandereycken},\ and\ \citenamefont {Verstraete}}]{haegeman2016unifying}%
  \BibitemOpen
  \bibfield  {author} {\bibinfo {author} {\bibfnamefont {J.}~\bibnamefont
  {Haegeman}}, \bibinfo {author} {\bibfnamefont {C.}~\bibnamefont {Lubich}},
  \bibinfo {author} {\bibfnamefont {I.}~\bibnamefont {Oseledets}}, \bibinfo
  {author} {\bibfnamefont {B.}~\bibnamefont {Vandereycken}}, \ and\ \bibinfo
  {author} {\bibfnamefont {F.}~\bibnamefont {Verstraete}},\ }\href
  {https://journals.aps.org/prb/abstract/10.1103/PhysRevB.94.165116} {\bibfield
   {journal} {\bibinfo  {journal} {Physical Review B}\ }\textbf {\bibinfo
  {volume} {94}},\ \bibinfo {pages} {165116} (\bibinfo {year}
  {2016})}\BibitemShut {NoStop}%
\bibitem [{\citenamefont {Blunden-Codd}\ \emph {et~al.}(2017)\citenamefont
  {Blunden-Codd}, \citenamefont {Bera}, \citenamefont {Bruognolo},
  \citenamefont {Linden}, \citenamefont {Chin}, \citenamefont {von Delft},
  \citenamefont {Nazir},\ and\ \citenamefont {Florens}}]{blunden2017anatomy}%
  \BibitemOpen
  \bibfield  {author} {\bibinfo {author} {\bibfnamefont {Z.}~\bibnamefont
  {Blunden-Codd}}, \bibinfo {author} {\bibfnamefont {S.}~\bibnamefont {Bera}},
  \bibinfo {author} {\bibfnamefont {B.}~\bibnamefont {Bruognolo}}, \bibinfo
  {author} {\bibfnamefont {N.-O.}\ \bibnamefont {Linden}}, \bibinfo {author}
  {\bibfnamefont {A.~W.}\ \bibnamefont {Chin}}, \bibinfo {author}
  {\bibfnamefont {J.}~\bibnamefont {von Delft}}, \bibinfo {author}
  {\bibfnamefont {A.}~\bibnamefont {Nazir}}, \ and\ \bibinfo {author}
  {\bibfnamefont {S.}~\bibnamefont {Florens}},\ }\href
  {https://journals.aps.org/prb/abstract/10.1103/PhysRevB.95.085104} {\bibfield
   {journal} {\bibinfo  {journal} {Physical Review B}\ }\textbf {\bibinfo
  {volume} {95}},\ \bibinfo {pages} {085104} (\bibinfo {year}
  {2017})}\BibitemShut {NoStop}%
\bibitem [{\citenamefont {Bulla}\ \emph {et~al.}(2003)\citenamefont {Bulla},
  \citenamefont {Tong},\ and\ \citenamefont {Vojta}}]{BullaPRL2003}%
  \BibitemOpen
  \bibfield  {author} {\bibinfo {author} {\bibfnamefont {R.}~\bibnamefont
  {Bulla}}, \bibinfo {author} {\bibfnamefont {N.-H.}\ \bibnamefont {Tong}}, \
  and\ \bibinfo {author} {\bibfnamefont {M.}~\bibnamefont {Vojta}},\ }\href
  {\doibase 10.1103/PhysRevLett.91.170601} {\bibfield  {journal} {\bibinfo
  {journal} {Phys. Rev. Lett.}\ }\textbf {\bibinfo {volume} {91}},\ \bibinfo
  {pages} {170601} (\bibinfo {year} {2003})}\BibitemShut {NoStop}%
\bibitem [{\citenamefont {Chin}\ \emph {et~al.}(2011)\citenamefont {Chin},
  \citenamefont {Prior}, \citenamefont {Huelga},\ and\ \citenamefont
  {Plenio}}]{ChinPRL2011}%
  \BibitemOpen
  \bibfield  {author} {\bibinfo {author} {\bibfnamefont {A.~W.}\ \bibnamefont
  {Chin}}, \bibinfo {author} {\bibfnamefont {J.}~\bibnamefont {Prior}},
  \bibinfo {author} {\bibfnamefont {S.~F.}\ \bibnamefont {Huelga}}, \ and\
  \bibinfo {author} {\bibfnamefont {M.~B.}\ \bibnamefont {Plenio}},\ }\href
  {\doibase 10.1103/PhysRevLett.107.160601} {\bibfield  {journal} {\bibinfo
  {journal} {Phys. Rev. Lett.}\ }\textbf {\bibinfo {volume} {107}},\ \bibinfo
  {pages} {160601} (\bibinfo {year} {2011})}\BibitemShut {NoStop}%
\bibitem [{\citenamefont {Weichselbaum}\ \emph {et~al.}(2009)\citenamefont
  {Weichselbaum}, \citenamefont {Verstraete}, \citenamefont {Schollw\"ock},
  \citenamefont {Cirac},\ and\ \citenamefont {von
  Delft}}]{WeichselbaumPRB2009}%
  \BibitemOpen
  \bibfield  {author} {\bibinfo {author} {\bibfnamefont {A.}~\bibnamefont
  {Weichselbaum}}, \bibinfo {author} {\bibfnamefont {F.}~\bibnamefont
  {Verstraete}}, \bibinfo {author} {\bibfnamefont {U.}~\bibnamefont
  {Schollw\"ock}}, \bibinfo {author} {\bibfnamefont {J.~I.}\ \bibnamefont
  {Cirac}}, \ and\ \bibinfo {author} {\bibfnamefont {J.}~\bibnamefont {von
  Delft}},\ }\href {\doibase 10.1103/PhysRevB.80.165117} {\bibfield  {journal}
  {\bibinfo  {journal} {Phys. Rev. B}\ }\textbf {\bibinfo {volume} {80}},\
  \bibinfo {pages} {165117} (\bibinfo {year} {2009})}\BibitemShut {NoStop}%
\bibitem [{\citenamefont {Winter}\ \emph {et~al.}(2009)\citenamefont {Winter},
  \citenamefont {Rieger}, \citenamefont {Vojta},\ and\ \citenamefont
  {Bulla}}]{WinterPRL2009}%
  \BibitemOpen
  \bibfield  {author} {\bibinfo {author} {\bibfnamefont {A.}~\bibnamefont
  {Winter}}, \bibinfo {author} {\bibfnamefont {H.}~\bibnamefont {Rieger}},
  \bibinfo {author} {\bibfnamefont {M.}~\bibnamefont {Vojta}}, \ and\ \bibinfo
  {author} {\bibfnamefont {R.}~\bibnamefont {Bulla}},\ }\href {\doibase
  10.1103/PhysRevLett.102.030601} {\bibfield  {journal} {\bibinfo  {journal}
  {Phys. Rev. Lett.}\ }\textbf {\bibinfo {volume} {102}},\ \bibinfo {pages}
  {030601} (\bibinfo {year} {2009})}\BibitemShut {NoStop}%
\bibitem [{\citenamefont {Nalbach}\ and\ \citenamefont
  {Thorwart}(2010{\natexlab{b}})}]{NalbachPRB2010}%
  \BibitemOpen
  \bibfield  {author} {\bibinfo {author} {\bibfnamefont {P.}~\bibnamefont
  {Nalbach}}\ and\ \bibinfo {author} {\bibfnamefont {M.}~\bibnamefont
  {Thorwart}},\ }\href {\doibase 10.1103/PhysRevB.81.054308} {\bibfield
  {journal} {\bibinfo  {journal} {Phys. Rev. B}\ }\textbf {\bibinfo {volume}
  {81}},\ \bibinfo {pages} {054308} (\bibinfo {year}
  {2010}{\natexlab{b}})}\BibitemShut {NoStop}%
\bibitem [{\citenamefont {Prior}\ \emph
  {et~al.}(2010{\natexlab{b}})\citenamefont {Prior}, \citenamefont {Chin},
  \citenamefont {Huelga},\ and\ \citenamefont {Plenio}}]{PriorPRL2010}%
  \BibitemOpen
  \bibfield  {author} {\bibinfo {author} {\bibfnamefont {J.}~\bibnamefont
  {Prior}}, \bibinfo {author} {\bibfnamefont {A.~W.}\ \bibnamefont {Chin}},
  \bibinfo {author} {\bibfnamefont {S.~F.}\ \bibnamefont {Huelga}}, \ and\
  \bibinfo {author} {\bibfnamefont {M.~B.}\ \bibnamefont {Plenio}},\ }\href
  {\doibase 10.1103/PhysRevLett.105.050404} {\bibfield  {journal} {\bibinfo
  {journal} {Phys. Rev. Lett.}\ }\textbf {\bibinfo {volume} {105}},\ \bibinfo
  {pages} {050404} (\bibinfo {year} {2010}{\natexlab{b}})}\BibitemShut
  {NoStop}%
\bibitem [{\citenamefont {Chin}\ \emph {et~al.}(2010)\citenamefont {Chin},
  \citenamefont {Rivas}, \citenamefont {Huelga},\ and\ \citenamefont
  {Plenio}}]{ChinJMP2010}%
  \BibitemOpen
  \bibfield  {author} {\bibinfo {author} {\bibfnamefont {A.~W.}\ \bibnamefont
  {Chin}}, \bibinfo {author} {\bibfnamefont {A.}~\bibnamefont {Rivas}},
  \bibinfo {author} {\bibfnamefont {S.~F.}\ \bibnamefont {Huelga}}, \ and\
  \bibinfo {author} {\bibfnamefont {M.~B.}\ \bibnamefont {Plenio}},\ }\href
  {http://aip.scitation.org/doi/full/10.1063/1.3490188} {\bibfield  {journal}
  {\bibinfo  {journal} {Journal of Mathematical Physics}\ }\textbf {\bibinfo
  {volume} {51}},\ \bibinfo {pages} {092109} (\bibinfo {year}
  {2010})}\BibitemShut {NoStop}%
\bibitem [{\citenamefont {Woods}\ \emph {et~al.}(2014)\citenamefont {Woods},
  \citenamefont {Groux}, \citenamefont {Chin}, \citenamefont {Huelga},\ and\
  \citenamefont {Plenio}}]{WoodsJMP2014}%
  \BibitemOpen
  \bibfield  {author} {\bibinfo {author} {\bibfnamefont {M.~P.}\ \bibnamefont
  {Woods}}, \bibinfo {author} {\bibfnamefont {R.}~\bibnamefont {Groux}},
  \bibinfo {author} {\bibfnamefont {A.~W.}\ \bibnamefont {Chin}}, \bibinfo
  {author} {\bibfnamefont {S.~F.}\ \bibnamefont {Huelga}}, \ and\ \bibinfo
  {author} {\bibfnamefont {M.~B.}\ \bibnamefont {Plenio}},\ }\href
  {http://scitation.aip.org/content/aip/journal/jmp/55/3/10.1063/1.4866769}
  {\bibfield  {journal} {\bibinfo  {journal} {Journal of Mathematical Physics}\
  }\textbf {\bibinfo {volume} {55}},\ \bibinfo {eid} {032101} (\bibinfo {year}
  {2014})}\BibitemShut {NoStop}%
\bibitem [{\citenamefont {Guo}\ \emph {et~al.}(2012)\citenamefont {Guo},
  \citenamefont {Weichselbaum}, \citenamefont {von Delft},\ and\ \citenamefont
  {Vojta}}]{guo2012critical}%
  \BibitemOpen
  \bibfield  {author} {\bibinfo {author} {\bibfnamefont {C.}~\bibnamefont
  {Guo}}, \bibinfo {author} {\bibfnamefont {A.}~\bibnamefont {Weichselbaum}},
  \bibinfo {author} {\bibfnamefont {J.}~\bibnamefont {von Delft}}, \ and\
  \bibinfo {author} {\bibfnamefont {M.}~\bibnamefont {Vojta}},\ }\href
  {https://journals.aps.org/prl/abstract/10.1103/PhysRevLett.108.160401}
  {\bibfield  {journal} {\bibinfo  {journal} {Physical review letters}\
  }\textbf {\bibinfo {volume} {108}},\ \bibinfo {pages} {160401} (\bibinfo
  {year} {2012})}\BibitemShut {NoStop}%
\bibitem [{\citenamefont {Kennes}\ \emph {et~al.}(2013)\citenamefont {Kennes},
  \citenamefont {Kashuba}, \citenamefont {Pletyukhov}, \citenamefont
  {Schoeller},\ and\ \citenamefont {Meden}}]{KennesPRL2013}%
  \BibitemOpen
  \bibfield  {author} {\bibinfo {author} {\bibfnamefont {D.~M.}\ \bibnamefont
  {Kennes}}, \bibinfo {author} {\bibfnamefont {O.}~\bibnamefont {Kashuba}},
  \bibinfo {author} {\bibfnamefont {M.}~\bibnamefont {Pletyukhov}}, \bibinfo
  {author} {\bibfnamefont {H.}~\bibnamefont {Schoeller}}, \ and\ \bibinfo
  {author} {\bibfnamefont {V.}~\bibnamefont {Meden}},\ }\href {\doibase
  10.1103/PhysRevLett.110.100405} {\bibfield  {journal} {\bibinfo  {journal}
  {Phys. Rev. Lett.}\ }\textbf {\bibinfo {volume} {110}},\ \bibinfo {pages}
  {100405} (\bibinfo {year} {2013})}\BibitemShut {NoStop}%
\bibitem [{\citenamefont {Strathearn}\ \emph {et~al.}(2017)\citenamefont
  {Strathearn}, \citenamefont {Lovett},\ and\ \citenamefont
  {Kirton}}]{StrathearnARXIV2017}%
  \BibitemOpen
  \bibfield  {author} {\bibinfo {author} {\bibfnamefont {A.}~\bibnamefont
  {Strathearn}}, \bibinfo {author} {\bibfnamefont {B.}~\bibnamefont {Lovett}},
  \ and\ \bibinfo {author} {\bibfnamefont {P.}~\bibnamefont {Kirton}},\ }\href
  {https://arxiv.org/abs/1704.04099} {\bibfield  {journal} {\bibinfo  {journal}
  {arXiv:}\ ,\ \bibinfo {pages} {1704.04099}} (\bibinfo {year}
  {2017})}\BibitemShut {NoStop}%
\bibitem [{\citenamefont {Blankenship}(2013)}]{blankenship2013molecular}%
  \BibitemOpen
  \bibfield  {author} {\bibinfo {author} {\bibfnamefont {R.~E.}\ \bibnamefont
  {Blankenship}},\ }\href@noop {} {\emph {\bibinfo {title} {Molecular
  mechanisms of photosynthesis}}}\ (\bibinfo  {publisher} {John Wiley \&
  Sons},\ \bibinfo {year} {2013})\BibitemShut {NoStop}%
\bibitem [{\citenamefont {Hu}\ \emph {et~al.}(2015)\citenamefont {Hu},
  \citenamefont {Hong}, \citenamefont {{Dean Smith}}, \citenamefont {Neusius},
  \citenamefont {Cheng},\ and\ \citenamefont {Smith}}]{Hu2015}%
  \BibitemOpen
  \bibfield  {author} {\bibinfo {author} {\bibfnamefont {X.}~\bibnamefont
  {Hu}}, \bibinfo {author} {\bibfnamefont {L.}~\bibnamefont {Hong}}, \bibinfo
  {author} {\bibfnamefont {M.}~\bibnamefont {{Dean Smith}}}, \bibinfo {author}
  {\bibfnamefont {T.}~\bibnamefont {Neusius}}, \bibinfo {author} {\bibfnamefont
  {X.}~\bibnamefont {Cheng}}, \ and\ \bibinfo {author} {\bibfnamefont {J.~C.}\
  \bibnamefont {Smith}},\ }\href {\doibase 10.1038/nphys3553} {\bibfield
  {journal} {\bibinfo  {journal} {Nature Physics}\ }\textbf {\bibinfo {volume}
  {12}},\ \bibinfo {pages} {171} (\bibinfo {year} {2015})}\BibitemShut
  {NoStop}%
\bibitem [{\citenamefont {Matyushov}(2013)}]{Matyushov2013}%
  \BibitemOpen
  \bibfield  {author} {\bibinfo {author} {\bibfnamefont {D.~V.}\ \bibnamefont
  {Matyushov}},\ }\href {\doibase 10.1063/1.4812788} {\bibfield  {journal}
  {\bibinfo  {journal} {Journal of Chemical Physics}\ }\textbf {\bibinfo
  {volume} {139}} (\bibinfo {year} {2013}),\ 10.1063/1.4812788}\BibitemShut
  {NoStop}%
\bibitem [{\citenamefont {Porras}\ \emph {et~al.}(2008)\citenamefont {Porras},
  \citenamefont {Marquardt}, \citenamefont {von Delft},\ and\ \citenamefont
  {Cirac}}]{PorrasPRA2008}%
  \BibitemOpen
  \bibfield  {author} {\bibinfo {author} {\bibfnamefont {D.}~\bibnamefont
  {Porras}}, \bibinfo {author} {\bibfnamefont {F.}~\bibnamefont {Marquardt}},
  \bibinfo {author} {\bibfnamefont {J.}~\bibnamefont {von Delft}}, \ and\
  \bibinfo {author} {\bibfnamefont {J.~I.}\ \bibnamefont {Cirac}},\ }\href
  {\doibase 10.1103/PhysRevA.78.010101} {\bibfield  {journal} {\bibinfo
  {journal} {Phys. Rev. A}\ }\textbf {\bibinfo {volume} {78}},\ \bibinfo
  {pages} {010101} (\bibinfo {year} {2008})}\BibitemShut {NoStop}%
\bibitem [{\citenamefont {Yu}\ \emph {et~al.}(2012)\citenamefont {Yu},
  \citenamefont {Tong}, \citenamefont {Xue}, \citenamefont {Wang},\ and\
  \citenamefont {Zhu}}]{YuScienceCHINA2012}%
  \BibitemOpen
  \bibfield  {author} {\bibinfo {author} {\bibfnamefont {L.}~\bibnamefont
  {Yu}}, \bibinfo {author} {\bibfnamefont {N.}~\bibnamefont {Tong}}, \bibinfo
  {author} {\bibfnamefont {Z.}~\bibnamefont {Xue}}, \bibinfo {author}
  {\bibfnamefont {Z.}~\bibnamefont {Wang}}, \ and\ \bibinfo {author}
  {\bibfnamefont {S.}~\bibnamefont {Zhu}},\ }\href {\doibase
  10.1007/s11433-012-4863-x} {\bibfield  {journal} {\bibinfo  {journal}
  {Science China Physics, Mechanics and Astronomy}\ }\textbf {\bibinfo {volume}
  {55}},\ \bibinfo {pages} {1557} (\bibinfo {year} {2012})}\BibitemShut
  {NoStop}%
\bibitem [{\citenamefont {Henriet}\ and\ \citenamefont
  {Le~Hur}(2016)}]{HenrietPRB2016}%
  \BibitemOpen
  \bibfield  {author} {\bibinfo {author} {\bibfnamefont {L.}~\bibnamefont
  {Henriet}}\ and\ \bibinfo {author} {\bibfnamefont {K.}~\bibnamefont
  {Le~Hur}},\ }\href {\doibase 10.1103/PhysRevB.93.064411} {\bibfield
  {journal} {\bibinfo  {journal} {Phys. Rev. B}\ }\textbf {\bibinfo {volume}
  {93}},\ \bibinfo {pages} {064411} (\bibinfo {year} {2016})}\BibitemShut
  {NoStop}%
\bibitem [{\citenamefont {Eckel}\ \emph {et~al.}(2009)\citenamefont {Eckel},
  \citenamefont {Reina},\ and\ \citenamefont {Thorwart}}]{eckel2009coherent}%
  \BibitemOpen
  \bibfield  {author} {\bibinfo {author} {\bibfnamefont {J.}~\bibnamefont
  {Eckel}}, \bibinfo {author} {\bibfnamefont {J.~H.}\ \bibnamefont {Reina}}, \
  and\ \bibinfo {author} {\bibfnamefont {M.}~\bibnamefont {Thorwart}},\ }\href
  {http://iopscience.iop.org/article/10.1088/1367-2630/11/8/085001/meta}
  {\bibfield  {journal} {\bibinfo  {journal} {New Journal of Physics}\ }\textbf
  {\bibinfo {volume} {11}},\ \bibinfo {pages} {085001} (\bibinfo {year}
  {2009})}\BibitemShut {NoStop}%
\end{thebibliography}%

\end{document}